\documentclass[conference]{IEEEtran}
\IEEEoverridecommandlockouts
\usepackage{cite}
\usepackage{amsmath,amssymb,amsfonts}
\usepackage{algorithmic}
\usepackage{graphicx}
\usepackage{textcomp}
\usepackage{xcolor}
\usepackage{cite} 
\usepackage{amsmath}
\usepackage[nolist]{acronym}
\usepackage{xfrac}
\usepackage{amssymb}
\usepackage{threeparttable}
\usepackage{siunitx} 
\usepackage[symbol]{footmisc}

\newcommand{\figref}[1]{Fig.~\ref{#1}}

\newcommand{\valunit}[2]{\ensuremath{{#1\,\mathrm{#2}}}}

\def\BibTeX{{\rm B\kern-.05em{\sc i\kern-.025em b}\kern-.08em
    T\kern-.1667em\lower.7ex\hbox{E}\kern-.125emX}}
\begin{document}
\begin{acronym}
    \acro{AI}{Artificial Intelligence}
    \acro{MAC}{Multipy-Accumulate}
    \acro{IMC}{In-Memory-Computing}
    \acro{LLM}{Large Language Model}
    \acro{DRAM}{dynamic Random Access Memory}
    \acro{ASIC}{Application-Specific Integrated Circuit}
    \acro{FPGA}{Field-Programmable Gate Array}
    \acro{PWM}{Pulse-Width Modulation}
    \acro{DAC}{Digital to Analog Converter}
    \acrodefplural{DAC}[DACs]{Digital to Analog Converters}
    \acro{ADC}{Analog to Digital Converter}
    \acrodefplural{ADC}[ADCs]{Analog to Digital Converters}
    \acro{IGZO}{Indium Gallium Zinc Oxide}
    \acro{ML}{Match Line}
    \acro{aCAM}{analog Contend Addressable Memory}
    \acrodefplural{aCAM}[aCAMs]{analog Contend Addressable Memories}
\end{acronym}
\title{Gain Cell-Based Analog Content Addressable Memory for Dynamic Associative tasks in AI}


\author{\IEEEauthorblockN{Paul-Philipp Manea\textsuperscript{1,3}, Nathan Leroux\textsuperscript{2}, Emre Neftci\textsuperscript{2,3}, John Paul Strachan\textsuperscript{1,3}}
\textsuperscript{1}PGI-14, Forschungszentrum Jülich, Aachen, Germany\\
\textsuperscript{2}PGI-15, Forschungszentrum Jülich, Aachen, Germany\\
\textsuperscript{3}RWTH Aachen University, Aachen, Germany\\
}

\def\retentiontime{$703\mu s$}

\maketitle

\begin{abstract}
Analog Content Addressable Memories (aCAMs) have proven useful for associative in-memory computing applications like Decision Trees, Finite State Machines, and Hyper-dimensional Computing. While non-volatile implementations using FeFETs and ReRAM devices offer speed, power, and area advantages, they suffer from slow write speeds and limited write cycles, making them less suitable for computations involving fully dynamic data patterns. To address these limitations, in this work, we propose a capacitor gain cell-based \ac{aCAM} designed for dynamic processing, where frequent memory updates are required. Our system compares analog input voltages to boundaries stored in capacitors, enabling efficient dynamic tasks. We demonstrate the application of \ac{aCAM} within transformer attention mechanisms by replacing the softmax-scaled dot-product similarity with \ac{aCAM} similarity, achieving competitive results.
Circuit simulations on a TSMC 28 nm node show promising performance in terms of energy efficiency, precision, and latency, making it well-suited for fast, dynamic AI applications.
\end{abstract}

\begin{IEEEkeywords}
In-memory computing, Capacitor Gain Cell, Dynamic Processing, Associative Computing, Non-Volatile Memory
\end{IEEEkeywords}

\section{Introduction}
\ac{IMC} has introduced a new paradigm in efficient computing for \ac{AI}, effectively addressing the memory bottleneck \cite{sebastian2020memory} driven by advancements in research of highly parallel, energy-efficient non-volatile memory technologies like ReRAM or FeFET. 

Gain cells, one of the emerging memory technologies, have been proven suitable for \ac{IMC} applications as well \cite{Wang2021,gou20232t1c} and have recently gained significant attention due to their CMOS compatibility, 3D integration capabilities, and suitability for dynamic tasks requiring frequent updates of matrix operands. Due to the oxide-based technologies, these devices have a retention in the order of multiple seconds making them suitable for real-world inference applications without the need for a memory refresh \cite{Shi3503, 10185398}.

The development of memristor-based \acp{aCAM}, which require fewer transistors compared to traditional SRAM-based designs \cite{Li2020}, has opened up new opportunities for various associative tasks in \ac{AI}, including Decision Trees, Random Forests, Finite State Machines, and Hyper-dimensional Computing \cite{pedretti2022differentiable, graves2022memory, kazemi2022achieving}. Furthermore, \acp{aCAM} have been implemented using various memory technologies, including FeFET \cite{yin2020fecam,LIU2024100218, 9106766} and Flash \cite{9502488}, as highlighted in Table~\ref{tab:aCAM_mem}. These \acp{aCAM} encode multi-level boundaries in memory, against which an analog input voltage is compared. If the input falls outside the boundary, a cell generates a mismatch current, otherwise not. A shared \ac{ML}, common to all cells performing a NOR operation, determines whether an entire word matches. In addition, a non-binary output can be yielded to measure the distance from the input query to the stored word keys, by measuring the mismatch timing or by adding all mismatch currents within the \ac{ML} using Kirchhoff's law \cite{pedretti2022differentiable}. This requires more complex readout circuits but offers greater capabilities.

Since many \ac{aCAM} systems are implemented using non-volatile technologies, they are subject to frequent non-idealities \cite{manea2023non, 9181475}. Specialized writing techniques are required to ensure correct values are reliably written to the device \cite{10405732}, which further extends programming times. Many non-volatile memories have a limited number of write cycles \cite{sebastian2020memory} compared to SRAM or DRAM, constraining their use in dynamic tasks that require frequent memory weight updates, even during inference, thereby restricting their applicability to static tasks. To address this limitation, in this work, we propose a capacitor gain cell-based \ac{aCAM} system, specifically designed for dynamic processing. To highlight its significance, we present an example task, where we integrate this \ac{aCAM} system within the attention mechanism of a transformer to compute the similarity between queries $Q$ and keys $K$ within one attention head.

The remainder of the paper is structured as follows: first, we introduce our dynamic \ac{aCAM} cell and describe its functionality, including a proposed macro architecture with peripheral circuitry. Next, we present our simulation results, focusing primarily on precision, speed, and energy consumption. Finally, we showcase application examples where dynamic processing is crucial, specifically in transformer attention inference.
\begin{table}[]
\caption{Comparison of \ac{aCAM} Implementations Using Various Non-Volatile Memory Technologies and the Proposed Work}
\label{tab:aCAM_mem}
\centering
\begin{threeparttable}
\setlength\tabcolsep{3pt}
\begin{tabular}{lllll}
                     & \textbf{FeFET}                           & \textbf{Flash}                                 & \textbf{ReRAM}                         & \textbf{This work}                \\ \hline
\textbf{Retention}   & \valunit{10}{Y} \cite{8630859}           & \valunit{10}{Y}                                & \valunit{5}{Y} \cite{lin2017retention} & \valunit{10^3}{s}\footnotemark[1] \\
\textbf{Write Speed} & \valunit{1}{\mu s} \cite{LIU2024100218}  & \valunit{10}{\mu s} \cite{sebastian2020memory} & \valunit{1}{\mu s} \cite{10405732}     & \valunit{20}{ns}                  \\
\textbf{Endurance}   & $10^5$ \cite{6242443}                    & $10^5$ \cite{sebastian2020memory}              & $10^6$ \cite{Jean2014Reliability}      & $10^{11}$\footnotemark[1]         \\
\textbf{Read Speed}  & \valunit{0.136}{ns} \cite{LIU2024100218} & \valunit{2.7}{ns} \cite{9502488}               & \valunit{220}{ps} \cite{Li2020}        & \valunit{6}{ns}                   \\
\textbf{No of Bits}  & $3$ \cite{9106766}                       & $3$ \cite{9502488}                             & $4$ \cite{manea2023non}                     & 3                                                         
\end{tabular}
\begin{tablenotes}  
    \item \footnotemark[1] IGZO write path \cite{9720596}.\\
    
\end{tablenotes}
\end{threeparttable}
\end{table}

\section{Dynamic aCAM Hardware}
\subsection{Capacitor gain cell-based dynamic \ac{aCAM} cell}
\label{sec:dynCAM_circ}
\begin{figure}[htb!]
    \centering
    \includegraphics[width=1\linewidth]{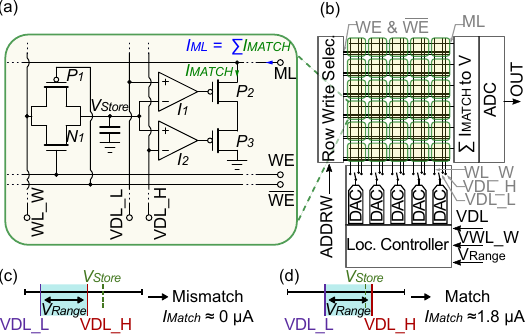}
    \caption{\textbf{(a)} Circuit configuration of the proposed dynamic \ac{aCAM} cell. \textbf{(b)} Architecture of the dynamic \ac{IMC} \ac{aCAM} macro including peripheral circuitry. \textbf{(c\&d)} Input voltage range within the dynamic search window for a given stored key voltage $V_{store}$ in both mismatch \textbf{(c)} and match \textbf{(d)} cases.}
    \label{fig:dyn_cam_circuit}
\end{figure}
The dynamic \ac{aCAM} demonstrator cell is implemented in TSMC's \valunit{28}{nm} technology and has been simulated in Cadence Virtuoso. The circuit for a single dynamic \ac{aCAM} cell of the proposed design is shown in \figref{fig:dyn_cam_circuit}~(a). For clarity, the circuit can be divided into four stages:
\begin{enumerate}
    \item A write enable stage, consisting of a transmission gate ($P_1$, $N_1$) that connects or disconnects the capacitor from the write word line (WL\_W).
    \item A storage capacitor $C_1$ that temporarily stores the reference voltage to which the input is compared.
    \item A comparator stage ($I_1$, $I_2$), where the input voltage is compared to the reference using two strong arm latched comparators: one for testing the lower and one for testing the upper bound. 
    \item The \ac{ML} stage ($P_2$, $P_3$), where the cell is connected to the common \ac{ML} via two PMOS transistors. This stage combines the results from all cells within a single word.
\end{enumerate}

\paragraph{Writing}
The writing process involves charging a capacitor with a multi-level voltage pulse lasting \valunit{20}{ns}, generated by a \ac{DAC}. The voltage pulse is gated to the target cell's capacitor via a write-enable (WE) signal controlling a transmission gate, allowing the capacitor to charge or discharge during the write phase. The stored voltage in the capacitor represents the reference for the subsequent comparison. As indicated in \figref{fig:dyn_cam_circuit}~(a), writing is performed row-wise by first applying the desired voltages to the WL\_W line and then enabling the WE lines.
\paragraph{Searching}
The inputs to the dynamic \ac{aCAM} are two signals, \textrm{VDL\_L} and \textrm{VDL\_H}. In contrast to previous approaches which stored the range within the cell and searched for a fixed voltage, our method applies the range directly at the input of the \ac{aCAM} and stores the search value within the cell, as shown in \figref{fig:dyn_cam_circuit}~(c\&d). To achieve this, a bias voltage $V_{Range}$ is added to or subtracted from the actual search voltage \textrm{VDL}, as shown in equations~\ref{eq:camVDL_H} and \ref{eq:camVDL_L}. 

\begin{equation}
\label{eq:camVDL_L}
\textrm{VDL\_L} = \textrm{VDL} - \sfrac{V_{Range}}{2} 
\end{equation}
\begin{equation}
\label{eq:camVDL_H}
\textrm{VDL\_H} = \textrm{VDL} + \sfrac{V_{Range}}{2} \text{.}
\end{equation}

The voltage  stored in the capacitor  \(V_{Store}\) is then compared to the two input voltages \textrm{VDL\_L} and \textrm{VDL\_H} using two strong-arm latched comparators, chosen for their fast response time. The first comparator determines if \(V_{Store}\) falls below the upper bound while the other checks if it exceeds the lower bound. The lower bound comparator $I_1$ outputs a high voltage if  \(V_{Store}\) is higher than \textrm{VDL\_L} and a low voltage if it is lower. The high bound comparator $I_2$ outputs a low voltage if \(V_{Store}\) is higher than \textrm{VDL\_H} and a high voltage if it is lower. The outputs of the comparators control two PMOS transistors acting as an AND gate. A match current, \(I_{MATCH}\), flows from the \ac{ML} to ground only when both comparator outputs are low, indicating that the input range overlaps with the stored reference. This behavior was chosen because the probability of a match is lower than that of a mismatch, allowing for energy savings. This function is graphically illustrated in Figure~\ref{fig:dyn_cam_circuit}~(a\&b), showing both match and mismatch cases, and corresponds to the logical formulation 
\begin{equation}
\label{eq:cam}
Match = \textrm{VDL\_L} < V_{Store} < \textrm{VDL\_H} \text{.}
\end{equation}
Similar to prior \ac{aCAM} designs \cite{Li2020}, where current flows only in the mismatch case, the P-channel devices ($P_2$, $P_3$) can be replaced by N-channel devices to achieve this functionality. 
\subsection{Macro system Architecture}
The macro architecture is shown in \figref{fig:dyn_cam_circuit}~(b), featuring an \(N \times M\) array along with the necessary peripheral circuitry. The \acp{DAC} serve two purposes: providing the write voltages for writing the capacitors using the WL\_W lines, and providing the search voltages. In the search phase, the same \acp{DAC} can be used to first pre-charge the lower bound input line with \textrm{VDL\_L} and then pre-charge the upper bound line with \textrm{VDL\_H}. This is possible because both signal lines only connect to transistor gates with low current leakage. The \acp{DAC} are controlled by a local controller that receives three input vectors of size $M$: the search voltage \textrm{VDL} for the queries, the write voltage for one row \textrm{WL\_W} for the keys as well as the range of the input search \textrm{RANGE}.

Since writing is performed row-wise, the target row is selected by the signal \textrm{ADDRW}, which controls a Row Write Selector that drives the \textrm{WE} lines. Simultaneously, the \acp{DAC} provide the voltages to be written to the capacitors. Notably, no verification step is required, offering a significant advantage over non-volatile memories.

All dynamic \ac{aCAM} cells forming a word are connected through a common NOR \ac{ML}. As shown in \figref{fig:dyn_cam_circuit}~(a), the first block converts the match currents to voltages, which are then sensed by \acp{ADC}. Since the output currents are integer multiples of \(I_{Match}\), corresponding to the number of matching cells, the \ac{ADC} bit precision must exceed \(\log_2(M)\) to accurately capture the current levels from each cell.

\section{Circuit Simulation Results}
In this section, we present the circuit simulation results obtained using the TSMC \valunit{28}{nm} process in Cadence Virtuoso. We begin with a precision analysis, followed by the operational speed evaluation, and conclude with energy consumption estimates.
\subsection{Precision}
\begin{figure}[htb!]
    \centering
    \includegraphics{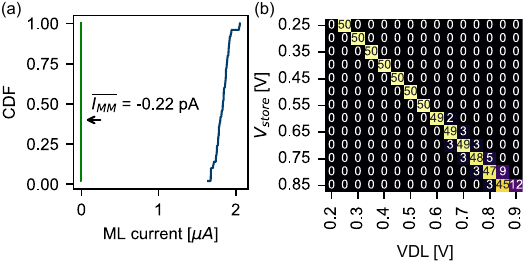}
    \caption{\textbf{(a)} Cumulative Distribution Function (CDF) from 50 Monte Carlo simulations, showing the distribution of \ac{ML} current amplitudes for match (blue) and mismatch (green) conditions in a dynamic \ac{aCAM} cell. Match condition: \(V_{Store} = 0.45 \, \textrm{V}\), \(\textrm{VDL} = 0.45 \, \textrm{V}\); Mismatch condition: \(V_{Store} = 0.45 \, \textrm{V}\), \(\textrm{VDL} = 0.40 \, \textrm{V}\). \textbf{(b)} Confusion matrix representing the results of 50 Monte Carlo simulations for all combinations of a search input and a stored reference. Diagonal entries indicate matching cases where \(V_{store} = \textrm{VDL}\). A match is determined by comparing the measured current to the threshold \(I_{match} = 350 \, \textrm{nA}\). The number of matches exceeding this threshold is counted, which is indicated by the numbers in each pixel.
}
    \label{fig:mc_data}
\end{figure}

Precision is a critical factor in ensuring that the \ac{aCAM} cell, operating in the analog domain, accurately returns a match when the input voltage range aligns with the stored voltage. This is achieved through Monte Carlo simulations, which account for process variations and mismatches. As a first step, we examine the current distribution for one match and one mismatch case, which is highlighted in \figref{fig:mc_data}~(b). Here the \ac{ML} currents are shown for 50 Monte Carlo simulations. The mean current in a mismatch scenario is \(\overline{I_{MM}} = 0.5\, \text{pA}\), in contrast to the match case, where the mean current is \(\overline{I_{M}} = 1.826\, \mu\text{A}\), resulting in a six orders of magnitude difference between match and mismatch, which ensures a reliable margin for distinguishing between them.

To evaluate the reliability across all parameter combinations, we perform again 50 Monte Carlo simulations with all possible combinations of  \( V_{store} \) and \textrm{VDL}, where in \figref{fig:mc_data}~(b) the number of detected Match currents is illustrated for each combination. True positives are prominently identified along the diagonal, where $\textrm{VDL} = V_{store}$. For larger values of $\textrm{VDL}$ and $V_{store}$ accuracy slightly drops. By analyzing the confusion matrix in \figref{fig:mc_data}~(b), we estimate the number of distinguishable levels to be 8, corresponding to 3 bits.
\subsection{Speed}
Dynamical processing requires to have both very fast read and write times. To evaluate the speed of or system we perform transient simulations using slow corners. To model the parasitics that occur in the wires of an array of size 128x128, we included a series resistance of $R_{ser}=1\Omega$ and a parasitic capacitance of $C_{par}=0.8 fF$ per array element.
\paragraph{Read speed}
A key characteristic of the strong-arm latched comparator design is that its speed depends on the input voltages. Lower input voltages cause the pre-charged nodes to discharge more slowly, leading to reduced speed. However, various biasing techniques exist to extend the dynamic range for smaller voltages \cite{8564870}, though these have not been considered in this work. To account for this, we evaluate the worst-case scenario, where \(\textrm{VDL} = V_{Store} = 250 \, \textrm{mV}\), defined as the lowest voltage within the dynamic search range. Under this condition, the achievable read time is \valunit{6}{ns}.
\paragraph{Write Speed}
To assess write speed, we conducted transient simulations with a parametric sweep of the initial voltage on the capacitor against the target voltage for the programming process. We then examined the required time for the capacitor to stabilize at the desired target voltage. Considering the worst-case scenario, which accounts for the longest write time within this parameter sweep, we note that a write time of \valunit{20}{ns} is required. Note that these results rely on our silicon CMOS simulations. 

\subsection{Energy Consumption}
In this section, we present the simulated energy consumption per cell. However, the energy consumption of peripheral circuitry is not included at this time.
\paragraph{Read/Search Energy}
The energy costs of a search operation are significantly dictated by whether a match occurs. Current flows through the Match Line (ML) only in the event of a match, leading to increased power consumption. We have calculated the power consumption to be \( 8.1 \, \text{fJ} \) for a match and \( 2.6 \, \text{fJ} \) for a mismatch. Assuming an equal probability distribution of input and stored values, the likelihood of a match is much lower (\( P(Match)=1/N \), where \(N\) is the number of intervals). This results in an estimated average energy consumption of \( 3.0 \, \text{fJ} \) per cell. Dividing this by the number of bits, the energy cost per search per bit approximates to \( \sim 1.0 \, \text{fJ}/\text{Search} / \text{Bit} \).   

\paragraph{Write Energy} 
The write energy in dynamic \ac{aCAM} is highly dependent on the voltage programmed into each cell. Estimating the mean energy over an equal distribution of write voltage we have calculated the write energy to be \( 4.8 \, \text{fJ} \).

\section{aCAM in Transformer Attention}
In this section, we demonstrate the versatility of our hardware by presenting an application example with dynamic requirements, where we integrate a software model of our dynamic \ac{aCAM} hardware into the transformer attention mechanisms. Before showcasing the results, we will also briefly introduce this task.

\begin{figure}[htb!]
    \centering
    \includegraphics{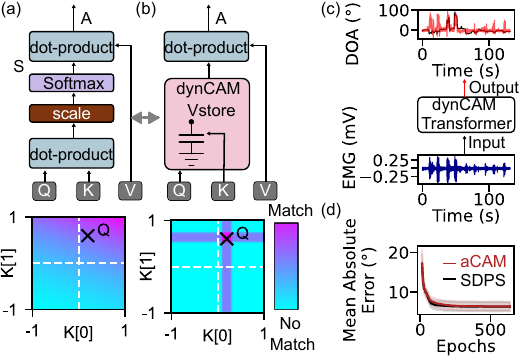}
    \caption{\textbf{(a)} Conventional attention head with scaled dot product similarity (SDPS). \textbf{(b)} proposed attention head using the dynamic aCAM for computing the similarity between $K$ and $Q$. \textbf{(c)} Electromyographic data processing task where the transformer predicts finger-joint angles. The predicted angles (degrees of actuation, DOA) and the target angles are respectively represented by red and black curves, while the input electromyographic signal is represented in blue. \textbf{(d)} Training results comparing conventional attention-based transformer with dynamic aCAM attention-based transformer.}
    \label{fig:transformer}
\end{figure}

Transformer networks \cite{vaswani2023attentionneed}, which have become the state-of-the-art for sequence processing, rely on the attention mechanism \cite{bahdanau_neural_2016} as shown in \figref{fig:transformer}~(a). In this mechanism, a similarity score is computed between the queries ($Q$) and keys ($K$) by applying a softmax to their scaled dot product. The resulting attention score ($S$) is then multiplied to the values ($V$). This type of similarity is favored due to its highly parallel computation on GPUs, making it ideal for large-scale models where efficiency is critical. However, various studies have shown that other similarity metrics, such as hierarchy-aware similarity measures based on hyperbolic entailment cones, inverse Euclidean or Fourier-based approaches, can enhance the performance of transformers  \cite{tseng2023coneheadshierarchyawareattention, bernhard2023alternativesscaleddotproduct, nguyen2022transformerfourierintegralattentions, shim2022similaritycontentbasedphoneticself,mccarter2023inversedistanceweightingattention}. 

In that study, we aim to demonstrate the use of dynamic \ac{aCAM} hardware to compute another type of similarity between queries ($Q$) and a matrix of keys ($K$), as illustrated in \figref{fig:transformer}~(b). To illustrate the differences between the similarity measures, we provide a visualization in a two-dimensional space of multiple key values ($K$) and a single query ($Q$). The colors in the plots represent the resulting similarity scores between the $Q$ and one $K$ vector corresponding to the location within the plane, emphasizing the improved selectivity of the \ac{aCAM} similarity compared to traditional scaled-dot product similarity.

Implementing attention within memory is challenging due to the large number of write operations needed. In previous research, we introduced gain cell-based in-memory computing hardware that utilizes a linear dot-product-based attention approach, which also relies on gain cell-based memories. This method offers a promising solution for addressing the dynamic nature of this task \cite{leroux2024attetnion}.

To evaluate the capacities of our \ac{aCAM} similarity-based attention, we implemented a PyTorch model of the dynamic \ac{aCAM} hardware \cite{Manea2023} and applied it to a bio-signal processing task, where dynamical processing is a necessity. For this study, we used  the NinaproDB8 dataset \cite{krasoulis_effect_2019}, where the goal is to process Electromyographic signals  representing muscle activity, measured on the forearm, in order to predict finger-joint angles. We used the same transformer model backbone and the same training procedure as in \cite{10388996}, replacing the softmax dot-product with the \ac{aCAM}-based attention. The inference process is described in see \figref{fig:transformer}~(c) where we also show an example of a temporal trace of predicted angles (degrees of actuation, DOA). In \figref{fig:transformer}~(d), we show the evolution over the training epochs of the mean absolute error averaged at different angles and time steps. Our transformer using \ac{aCAM} similarity achieves a mean absolute error of 6.35°,  comparable to the error of the transformer based on the conventional scaled dot-product similarity (SDPS) of 6.33°, thus demonstrating the trainability of the differential \ac{aCAM}, and their potential to compute attention scores.

\section{Conclusion}
While \ac{aCAM} systems are a useful \ac{IMC} tool for enhancing applications requiring associative memory by offering ultra-fast and energy-efficient similarity computations, current designs rely on non-volatile devices, which come with high write costs. As a result, they are not well-suited for dynamic in-memory computing tasks.

To address this issue, we introduce a dynamic \ac{aCAM} design based on gain cell technology, leveraging advancements in 2D oxide-based semiconductors that offer long retention times and CMOS compatibility. Our design supports multi-level storage, estimated to 8 levels per cell (3 Bits) without requiring a program and verification scheme common to most non-volatile memory technologies. We highlight the low write energy consumption of these devices, requiring only \valunit{4.8}{fJ} per cell, and a duration of \valunit{20}{ns}. A search operation can be performed with an energy consumption of \valunit{3.0}{fJ} per cell, completing in just \valunit{6}{ns}. We showcase one example of dynamic associative processing for our system, which is an alternative method for computing similarity scores within a transformer's attention mechanism.

We introduced a new building block to the field of \ac{IMC}, expanding its application by incorporating dynamic associative memory. This is particularly useful in scenarios where not only the queries, representing the search vector, are dynamic, but also the keys, which are represented by the memory weights are updated frequently. In conclusion, our hardware dynamical associative memory implementation is a new building block paving the way toward more efficient and versatile \ac{IMC} systems.

\section*{Acknowledgments}
This work was supported in part by the Federal Ministry of Education and Research (BMBF, Germany) in the project NEUROTEC II (Project number: 16ME0398K). This work is based on the Jülich Aachen Research Alliance (JARA-FIT) at Forschungszentrum Jülich GmbH, Jülich, Germany.
\newpage

\begin{thebibliography}{10}
\providecommand{\url}[1]{#1}
\csname url@samestyle\endcsname
\providecommand{\newblock}{\relax}
\providecommand{\bibinfo}[2]{#2}
\providecommand{\BIBentrySTDinterwordspacing}{\spaceskip=0pt\relax}
\providecommand{\BIBentryALTinterwordstretchfactor}{4}
\providecommand{\BIBentryALTinterwordspacing}{\spaceskip=\fontdimen2\font plus
\BIBentryALTinterwordstretchfactor\fontdimen3\font minus
  \fontdimen4\font\relax}
\providecommand{\BIBforeignlanguage}[2]{{%
\expandafter\ifx\csname l@#1\endcsname\relax
\typeout{** WARNING: IEEEtran.bst: No hyphenation pattern has been}%
\typeout{** loaded for the language `#1'. Using the pattern for}%
\typeout{** the default language instead.}%
\else
\language=\csname l@#1\endcsname
\fi
#2}}
\providecommand{\BIBdecl}{\relax}
\BIBdecl

\bibitem{sebastian2020memory}
A.~Sebastian, M.~Le~Gallo, R.~Khaddam-Aljameh, and E.~Eleftheriou, ``Memory
  devices and applications for in-memory computing,'' \emph{Nature
  nanotechnology}, vol.~15, no.~7, pp. 529--544, 2020.

\bibitem{Wang2021}
Y.~Wang, H.~Tang, Y.~Xie, X.~Chen, S.~Ma, Z.~Sun, Q.~Sun, L.~Chen, H.~Zhu,
  J.~Wan, Z.~Xu, D.~W. Zhang, P.~Zhou, and W.~Bao, ``An in-memory computing
  architecture based on two-dimensional semiconductors for multiply-accumulate
  operations,'' \emph{Nature Communications}, vol.~12, no.~1, jun 2021.

\bibitem{gou20232t1c}
S.~Gou, Y.~Wang, X.~Dong, Z.~Xu, X.~Wang, Q.~Sun, Y.~Xie, P.~Zhou, and W.~Bao,
  ``2t1c dram based on semiconducting mos2 and semimetallic graphene for
  in-memory computing,'' \emph{National Science Open}, vol.~2, no.~4, p.
  20220071, 2023.

\bibitem{Shi3503}
M.~Shi, Y.~Su, J.~Tang, Y.~Li, Y.~Du, R.~An, J.~Li, Y.~Li, J.~Yao, R.~Hu,
  Y.~He, Y.~Xi, Q.~Li, S.~Qiu, Q.~Zhang, L.~Pan, B.~Gao, H.~Qian, and H.~Wu,
  ``Counteractive coupling igzo/cnt hybrid 2t0c dram accelerating rram-based
  computing-in-memory via monolithic 3d integration for edge ai.''\hskip 1em
  plus 0.5em minus 0.4em\relax IEEE, 3503.

\bibitem{10185398}
A.~Belmonte, S.~Kundu, S.~Subhechha, A.~Chasin, N.~Rassoul, H.~Dekkers,
  H.~Puliyalil, F.~Seidel, P.~Carolan, R.~Delhougne, and G.~S. Kar, ``Lowest
  ioff \(<\) 3×10-21 a/\textmu m in capacitorless dram achieved by reactive
  ion etch of igzo-tft,'' in \emph{2023 IEEE Symposium on VLSI Technology and
  Circuits (VLSI Technology and Circuits)}, 2023, pp. 1--2.

\bibitem{Li2020}
C.~Li, C.~E. Graves, X.~Sheng, D.~Miller, M.~Foltin, G.~Pedretti, and J.~P.
  Strachan, ``Analog content-addressable memories with memristors,''
  \emph{Nature communications}, vol.~11, no.~1, p. 1638, 2020.

\bibitem{pedretti2022differentiable}
G.~Pedretti, C.~E. Graves, T.~Van~Vaerenbergh, S.~Serebryakov, M.~Foltin,
  X.~Sheng, R.~Mao, C.~Li, and J.~P. Strachan, ``Differentiable content
  addressable memory with memristors,'' \emph{Advanced electronic materials},
  vol.~8, no.~8, p. 2101198, 2022.

\bibitem{graves2022memory}
C.~E. Graves, C.~Li, G.~Pedretti, and J.~P. Strachan, ``In-memory computing
  with non-volatile memristor cam circuits,'' in \emph{Memristor Computing
  Systems}.\hskip 1em plus 0.5em minus 0.4em\relax Springer, 2022, pp.
  105--139.

\bibitem{kazemi2022achieving}
A.~Kazemi, F.~M{\"u}ller, M.~M. Sharifi, H.~Errahmouni, G.~Gerlach,
  T.~K{\"a}mpfe, M.~Imani, X.~S. Hu, and M.~Niemier, ``Achieving
  software-equivalent accuracy for hyperdimensional computing with
  ferroelectric-based in-memory computing,'' \emph{Scientific reports},
  vol.~12, no.~1, p. 19201, 2022.

\bibitem{yin2020fecam}
X.~Yin, C.~Li, Q.~Huang, L.~Zhang, M.~Niemier, X.~S. Hu, C.~Zhuo, and K.~Ni,
  ``Fecam: A universal compact digital and analog content addressable memory
  using ferroelectric,'' \emph{IEEE Transactions on Electron Devices}, vol.~67,
  no.~7, pp. 2785--2792, 2020.

\bibitem{LIU2024100218}
\BIBentryALTinterwordspacing
X.~Liu, K.~Katti, Y.~He, P.~Jacob, C.~Richter, U.~Schroeder, S.~Kurinec,
  P.~Chaudhari, and D.~Jariwala, ``Analog content-addressable memory from
  complementary fefets,'' \emph{Device}, vol.~2, no.~2, p. 100218, 2024.
  [Online]. Available:
  \url{https://www.sciencedirect.com/science/article/pii/S2666998623003587}
\BIBentrySTDinterwordspacing

\bibitem{9106766}
X.~Yin, C.~Li, Q.~Huang, L.~Zhang, M.~Niemier, X.~S. Hu, C.~Zhuo, and K.~Ni,
  ``Fecam: A universal compact digital and analog content addressable memory
  using ferroelectric,'' \emph{IEEE Transactions on Electron Devices}, vol.~67,
  no.~7, pp. 2785--2792, 2020.

\bibitem{9502488}
A.~Kazemi, S.~Sahay, A.~Saxena, M.~M. Sharifi, M.~Niemier, and X.~S. Hu, ``A
  flash-based multi-bit content-addressable memory with euclidean squared
  distance,'' in \emph{2021 IEEE/ACM International Symposium on Low Power
  Electronics and Design (ISLPED)}, 2021, pp. 1--6.

\bibitem{manea2023non}
P.-P. Manea, C.~Sudarshan, F.~C{\"u}ppers, and J.~P. Strachan, ``Non-idealities
  and design solutions for analog memristor-based content-addressable
  memories,'' in \emph{Proceedings of the 18th ACM International Symposium on
  Nanoscale Architectures}, 2023, pp. 1--6.

\bibitem{9181475}
C.~Bengel, A.~Siemon, F.~Cüppers, S.~Hoffmann-Eifert, A.~Hardtdegen, M.~von
  Witzleben, L.~Hellmich, R.~Waser, and S.~Menzel, ``Variability-aware modeling
  of filamentary oxide-based bipolar resistive switching cells using spice
  level compact models,'' \emph{IEEE Transactions on Circuits and Systems I:
  Regular Papers}, vol.~67, no.~12, pp. 4618--4630, 2020.

\bibitem{10405732}
J.~Yu, P.-P. Manea, S.~Ameli, M.~Hizzani, A.~Eldebiky, and J.~P. Strachan,
  ``Analog feedback-controlled memristor programming circuit for analog content
  addressable memory,'' in \emph{2023 IEEE International Conference on
  Metrology for eXtended Reality, Artificial Intelligence and Neural
  Engineering (MetroXRAINE)}, 2023, pp. 983--988.

\bibitem{8630859}
K.-T. Chen, H.-Y. Chen, C.-Y. Liao, G.-Y. Siang, C.~Lo, M.-H. Liao, K.-S. Li,
  S.~T. Chang, and M.~H. Lee, ``Non-volatile ferroelectric fets using 5-nm
  hf0.5zr0.5o2 with high data retention and read endurance for 1t memory
  applications,'' \emph{IEEE Electron Device Letters}, vol.~40, no.~3, pp.
  399--402, 2019.

\bibitem{lin2017retention}
Y.-D. Lin, P.-S. Chen, H.-Y. Lee, Y.-S. Chen, S.~Z. Rahaman, K.-H. Tsai, C.-H.
  Hsu, W.-S. Chen, P.-H. Wang, Y.-C. King \emph{et~al.}, ``Retention model of
  tao/hfo x and tao/alo x rram with self-rectifying switch characteristics,''
  \emph{Nanoscale research letters}, vol.~12, pp. 1--6, 2017.

\bibitem{6242443}
J.~Müller, E.~Yurchuk, T.~Schlösser, J.~Paul, R.~Hoffmann, S.~Müller,
  D.~Martin, S.~Slesazeck, P.~Polakowski, J.~Sundqvist, M.~Czernohorsky,
  K.~Seidel, P.~Kücher, R.~Boschke, M.~Trentzsch, K.~Gebauer, U.~Schröder,
  and T.~Mikolajick, ``Ferroelectricity in hfo2 enables nonvolatile data
  storage in 28 nm hkmg,'' in \emph{2012 Symposium on VLSI Technology (VLSIT)},
  2012, pp. 25--26.

\bibitem{Jean2014Reliability}
Y.-S. Jean, M.~Fazio, M.~Amrbar, M.~White, and D.~Sheldon, ``Reliability
  characterization of a commercial taox-based reram,'' \emph{2014 IEEE
  International Integrated Reliability Workshop Final Report (IIRW)}, pp.
  131--134, 2014.

\bibitem{9720596}
A.~Belmonte, H.~Oh, S.~Subhechha, N.~Rassoul, H.~Hody, H.~Dekkers,
  R.~Delhougne, L.~Ricotti, K.~Banerjee, A.~Chasin, M.~J. van Setten,
  H.~Puliyalil, M.~Pak, L.~Teugels, D.~Tsvetanova, K.~Vandersmissen, S.~Kundu,
  J.~Heijlen, D.~Batuk, J.~Geypen, L.~Goux, and G.~S. Kar, ``Tailoring igzo-tft
  architecture for capacitorless dram, demonstrating > 103s retention, >1011
  cycles endurance and lg scalability down to 14nm,'' in \emph{2021 IEEE
  International Electron Devices Meeting (IEDM)}, 2021, pp. 10.6.1--10.6.4.

\bibitem{8564870}
J.~Lan, Y.~Chen, Z.~Ni, F.~Ye, and J.~Ren, ``High speed, low power dynamic
  comparator with dynamic biasing in pre-amplifier phase,'' in \emph{2018 14th
  IEEE International Conference on Solid-State and Integrated Circuit
  Technology (ICSICT)}, 2018, pp. 1--3.

\bibitem{vaswani2023attentionneed}
\BIBentryALTinterwordspacing
A.~Vaswani, N.~Shazeer, N.~Parmar, J.~Uszkoreit, L.~Jones, A.~N. Gomez,
  L.~Kaiser, and I.~Polosukhin, ``Attention is all you need,'' 2023. [Online].
  Available: \url{https://arxiv.org/abs/1706.03762}
\BIBentrySTDinterwordspacing

\bibitem{bahdanau_neural_2016}
\BIBentryALTinterwordspacing
D.~Bahdanau, K.~Cho, and Y.~Bengio, ``\BIBforeignlanguage{en}{Neural {Machine}
  {Translation} by {Jointly} {Learning} to {Align} and {Translate}},'' May
  2016, arXiv:1409.0473 [cs, stat]. [Online]. Available:
  \url{http://arxiv.org/abs/1409.0473}
\BIBentrySTDinterwordspacing

\bibitem{tseng2023coneheadshierarchyawareattention}
\BIBentryALTinterwordspacing
A.~Tseng, T.~Yu, T.~J.~B. Liu, and C.~D. Sa, ``Coneheads: Hierarchy aware
  attention,'' 2023. [Online]. Available:
  \url{https://arxiv.org/abs/2306.00392}
\BIBentrySTDinterwordspacing

\bibitem{bernhard2023alternativesscaleddotproduct}
\BIBentryALTinterwordspacing
J.~Bernhard, ``Alternatives to the scaled dot product for attention in the
  transformer neural network architecture,'' 2023. [Online]. Available:
  \url{https://arxiv.org/abs/2311.09406}
\BIBentrySTDinterwordspacing

\bibitem{nguyen2022transformerfourierintegralattentions}
\BIBentryALTinterwordspacing
T.~Nguyen, M.~Pham, T.~Nguyen, K.~Nguyen, S.~J. Osher, and N.~Ho, ``Transformer
  with fourier integral attentions,'' 2022. [Online]. Available:
  \url{https://arxiv.org/abs/2206.00206}
\BIBentrySTDinterwordspacing

\bibitem{shim2022similaritycontentbasedphoneticself}
\BIBentryALTinterwordspacing
K.~Shim and W.~Sung, ``Similarity and content-based phonetic self attention for
  speech recognition,'' 2022. [Online]. Available:
  \url{https://arxiv.org/abs/2203.10252}
\BIBentrySTDinterwordspacing

\bibitem{mccarter2023inversedistanceweightingattention}
\BIBentryALTinterwordspacing
C.~McCarter, ``Inverse distance weighting attention,'' 2023. [Online].
  Available: \url{https://arxiv.org/abs/2310.18805}
\BIBentrySTDinterwordspacing

\bibitem{leroux2024attetnion}
\BIBentryALTinterwordspacing
N.~Leroux, P.-P. Manea, C.~Sudarshan, J.~Finkbeiner, S.~Siegel, J.~P. Strachan,
  and E.~Neftci, ``Analog in-memory computing attention mechanism for fast and
  energy-efficient large language models,'' 2024. [Online]. Available:
  \url{https://arxiv.org/abs/2409.19315}
\BIBentrySTDinterwordspacing

\bibitem{Manea2023}
P.-P. Manea, ``torchcam,'' \url{https://iffgit.fz-juelich.de/manea/torchcam},
  2023.

\bibitem{krasoulis_effect_2019}
\BIBentryALTinterwordspacing
A.~Krasoulis, S.~Vijayakumar, and K.~Nazarpour, ``Effect of {User} {Practice}
  on {Prosthetic} {Finger} {Control} {With} an {Intuitive} {Myoelectric}
  {Decoder},'' \emph{Frontiers in Neuroscience}, vol.~13, 2019. [Online].
  Available:
  \url{https://www.frontiersin.org/articles/10.3389/fnins.2019.00891}
\BIBentrySTDinterwordspacing

\bibitem{10388996}
N.~Leroux, J.~Finkbeiner, and E.~Neftci, ``Online transformers with spiking
  neurons for fast prosthetic hand control,'' in \emph{2023 IEEE Biomedical
  Circuits and Systems Conference (BioCAS)}, 2023, pp. 1--6.

\end{thebibliography}

\end{document}